\documentclass[useAMS,usenatbib,usegraphicx]{mn2e}

\def\sles{\lower2pt\hbox{$\buildrel {\scriptstyle <}
   \over {\scriptstyle\sim}$}}

\def\sgreat{\lower2pt\hbox{$\buildrel {\scriptstyle >}
   \over {\scriptstyle\sim}$}}

\title[The pulsar synchrotron: coherent radio emission]
{The pulsar synchrotron: coherent radio emission}
\author[I. Contopoulos]{I. Contopoulos\thanks{E-mail:
icontop@academyofathens.gr}\\
Research Center for Astronomy, Academy of Athens, 4
Soranou-Efessiou Str., Athens 11527, Greece}
\begin{document}

\date{Accepted 2009 March 3.  Received 2009 February 10; in original
form 2008 November 11}

\pagerange{\pageref{firstpage}--\pageref{lastpage}} \pubyear{2009}

\maketitle

\label{firstpage}

\begin{abstract}
We propose a simple physical picture for the generation of
coherent radio emission in the axisymmetric pulsar magnetosphere
that is quite different from the canonical paradigm of radio
emission coming from the magnetic polar caps. In this first paper
we consider only the axisymmetric case of an aligned rotator. Our
picture capitalizes on an important element of the MHD
representation of the magnetosphere, namely the separatrix between
the corotating closed-line region (the `dead zone') and the open
field lines that originate in the polar caps. Along the separatrix
flows the return current that corresponds to the main
magnetospheric electric current emanating from the polar caps.
Across the separatrix, both the toroidal and poloidal components
of the magnetic field change discontinuously. The poloidal
component discontinuity requires the presence of a significant
annular electric current which has up to now been unaccounted for.
We estimate the position and thickness of this annular current at
the tip of the closed line region, and show that it consists of
electrons (positrons) corotating with Lorentz factors on the order
of $10^5$, emitting incoherent synchrotron radiation that peaks in
the hard X-rays. These particles stay in the region of highest
annular current close to the equator for a path-length of the
order of one meter. We propose that, at wavelengths comparable to
that path-length, the particles emit coherent radiation, with
radiated power proportional to $N^2$, where $N$ is the population
of particles in the above path-length. We calculate the total
radio power in this wavelength regime and its scaling with pulsar
period and stellar magnetic field and show that it is consistent
with estimates of radio luminosity based on observations.
\end{abstract}

\begin{keywords}
pulsars -- radio: stars.
\end{keywords}

\section{Introduction}

The physical mechanism of pulsar radio emission remains still
poorly understood. Observational evidence (e.g. Kramer, Wex \&
Wielebinski~2000) suggests that it must be due to some coherent
radiation mechanism. Unfortunately, neither the exact mechanism
(curvature radiation, free electron maser emission, excited plasma
waves, etc.), nor its origin (inner or outer magnetosphere) are
clear. It is nevertheless widely accepted that radio emission
originates within a few stellar radii from the stellar surface,
and is due to relativistic plasma outflowing from the two stellar
magnetic poles (e.g. Kramer {\em et al.}~1997).

We here would like to propose a new physical picture for pulsar
radio emission that capitalizes on what we have learned about the
pulsar magnetosphere when studied under the assumption of
force-free ideal magnetohydrodynamics. The first self-consistent
solution of the axisymmetric problem (Contopoulos, Kazanas \&
Fendt~1999, hereafter CKF) demonstrated that the pulsar
magnetosphere consists of a closed line corotating region (the so
called `dead zone') that extends up to the `light cylinder' at
distances
\begin{equation}
r_{lc}\equiv c/\Omega_*=1.6\times 10^8\mbox{cm}\left(
\frac{P_*}{33\mbox{ms}}\right)
\end{equation}
from the rotation axis ($\Omega_*$ is the angular velocity and
$P_*\equiv 2\pi/\Omega_*$ the period of the rotating neutron
star), and two open line regions that originate around the two
stellar magnetic poles (the so called `polar caps'). From each
polar cap outflows the main magnetospheric current, which in the
case of the aligned rotator may be estimated as
\begin{equation}
I=\mp 1.23 \frac{B_* r_*^3 \Omega_*^2}{4c}\ . \label{I}
\end{equation}
Here, $B_*\sim 10^{13}\mbox{G}$ and $r_*\sim 10^6\mbox{cm}$ are
the stellar polar magnetic field and radius respectively
(Gruzinov~2004; Contopoulos~2005). The $\mp$ sign in eq.~(\ref{I})
depends on whether the magnetic axis is aligned or counter-aligned
with the axis of rotation (our results will heretofore refer to
the aligned case with the corresponding result for the
counter-aligned case in parentheses). CKF established that the
return electric current flows along the separatrix between open
and closed field lines. In fact, they self-consistently determined
the distribution of electric current as well as the distribution
of electric space-charge everywhere in the magnetosphere. We would
like to point out here that magnetohydrodynamics does not deal
with the issue of how this is materialized with actual
distributions and flows of charged particles (electrons and
positrons). This is not really a problem for the largest part of
the magnetosphere since it is known that the number density of the
electron-positron plasma that fills the magnetosphere exceeds the
above space-charge by a huge multiplicity factor (on the order of
$10^3-10^4$; e.g. Arons~1983), and all that is needed in order to
obtain the distributions required by CKF is a {\em slight
difference in the velocities of electrons and positrons}. As we
will now argue, this may not be true in the separatrix between
open and closed field lines.

\section{The pulsar synchrotron}

We will discuss here only the axisymmetic case (we plan to address
the 3D problem in Paper~II of this series based on the numerical
results of Kalapotharakos \& Contopoulos~2009). Inside the closed
line region, $B_\phi=0$, and particles at its tip corotate with
Lorentz factor $\gamma$ equal to
\begin{equation}
\gamma \equiv \frac{1}{\sqrt{1-(R\Omega_*/c)^2}}\approx
\frac{1}{\sqrt{2(r_{lc}-R)/r_{lc}}}\ ,\label{gammaS}
\end{equation}
where, $R\approx r_{lc}$ is the (cylindrical) distance from the
rotation axis. Let us heretofore denote quantities at the
inner/outer surface of the separatrix with $\mp$ indices
respectively. The toroidal magnetic field $B_\phi^-=0$, and
\begin{equation}
B_\phi^+ = \frac{2I}{cR}\ .\label{Bphi}
\end{equation}
Uzdensky~(2003) and others pointed out that, as the tip of the
dead zone approaches the light cylinder, the poloidal magnetic
field $B_p^-$ diverges as
\begin{equation}
\frac{B_p^-}{B_\phi^+}=\gamma\ \label{BpBphi}
\end{equation}
and $B_p^+=0$ (the Lorenz invariant $B^2-E^2$ is continuous across
the separatrix, and $E=R\Omega_*B_p/c$; see Ffig.~1; see also
fig.~4 of Kalapotharakos \& Contopoulos~2009, fig.~11 of
Timokhin~2006, and fig.~1 of Spitkovsky~2006).

Note that particles at the tip of the dead zone corotate with the
star due to the combined action of the magnetospheric electric and
magnetic fields. During that motion, they deviate around their
corotation radius within distances on the order of the gyroradius
\begin{eqnarray}
d & \approx & \frac{\gamma m_e c^2}{eB_p^-} =\frac{m_e c^2}{e
B_\phi^+}
\equiv \frac{c}{\omega_B}\nonumber \\
& \approx & 10^{-3} \mbox{cm}
\left(\frac{B_*}{10^{13}\mbox{G}}\right)^{-1} \left(
\frac{P_*}{33\mbox{ms}} \right)^3\ .\label{d1}
\end{eqnarray}
Here, $m_e$, $e$ are the electron mass and charge respectively,
and $\omega_B\equiv eB_p^-/\gamma m_e c$ is the gyration
frequency. For a `cold' particle distribution, {\em this distance
dictates also the thickness of the separatrix layer}. We will
assume that the closed line region extends as close to the light
cylinder as is allowed by the equations of motion\footnote{That
is, we will ignore here previous solutions (Contopoulos~2005;
Timokhin~2006) which suggested that the dead zone may end at any
distance inside the light cylinder.}, namely that
\begin{eqnarray}
d & \approx & r_{lc}-R=\frac{r_{lc}}{2\gamma^2}\ .\label{d2}
\end{eqnarray}
Eqs.~(\ref{d1}) \& (\ref{d2}) yield
\begin{eqnarray}
\gamma & = & \sqrt{\frac{r_{lc}}{2d}} \equiv
\sqrt{\frac{\omega_B}{2\Omega_*}}
\nonumber \\
& \approx & 3\times 10^5
\left(\frac{B_*}{10^{13}\mbox{G}}\right)^{1/2}
\left(\frac{P_*}{33\mbox{ms}}\right)^{-1}\ .\label{gamma}
\end{eqnarray}

One interesting feature of the separatrix is that it is negatively
(positively) charged with surface charge density
\begin{equation}
\sigma \equiv \frac{E^+ - E^-}{4\pi}\approx \mp
\frac{B_p^-}{4\pi}\ . \label{sigma}
\end{equation}
We can express the number density $n$ of charge carriers in the
separatrix layer as
\begin{equation}
n\equiv \kappa \frac{\sigma}{ed}=\kappa
\frac{(B_p^-)^2}{4\pi\gamma m_e c^2}\ .\label{kappa}
\end{equation}
Obviously, {\em the multiplicity factor $\kappa$ must be on the
order of unity for the force-free assumption to be valid}. Note
that at the tip of the dead zone, the Goldreich-Julian number
density $n_{GJ}$ is equal to
\begin{equation}
n_{GJ}\equiv \frac{-{\bf B}\cdot {\bf \Omega}_*}{2\pi e
c[1-(R/r_{lc})^2]} \approx \frac{B_p^-}{4\pi e
d}=\frac{(B_p^-)^2}{4\pi\gamma m_e c^2}\label{nGJ}
\end{equation}
(Goldreich \& Julian~1969, CKF), and therefore,
\begin{eqnarray}
n & \approx & n_{GJ}\nonumber\\
 & = & 6\times
10^{22}\mbox{cm}^{-3}\left(\frac{B_*}{10^{13}\mbox{G}}\right)^{5/2}
\left( \frac{P_*}{33\mbox{ms}} \right)^{-7}\ .
\end{eqnarray}
We have shown here that the number density of charge carriers at
the tip of the separatrix must be close to the minimum number
density needed to support the surface charge density required by
eq.~(\ref{sigma}). These particles corotate with the star at
speeds very close to the speed of light. If we now apply
Amp\`{e}re's law
\begin{equation}
B_p^- = \frac{4\pi}{c}\frac{{\rm d}I_\phi}{{\rm d}l} \approx 4\pi
\sigma c\ ,\label{BS}
\end{equation}
and if we assume that particles at the tip of the separatrix move
with velocities $\vec{\beta} c$, then eqs.~(\ref{BS}) \&
(\ref{Bphi}) imply that
\begin{equation}
\frac{\beta_p}{\beta_\phi}\approx \beta_p \sim
\frac{B_\phi^+}{B_p^-}=\frac{1}{\gamma}\ .\label{beta}
\end{equation}
Note that the separatrix is negatively (positively) charged, and
therefore, $\vec{\beta}$ points along the direction of stellar
rotation and towards the star. In other words, {\em electrons
(positrons) in the separatrix follow a spiral motion as they move
inwards away from the tip of the closed line region, with very
small pitch angles on the order of $1/\gamma$} (fig.~2).

CKF (and for that matter all subsequent investigations) failed to
emphasize the significance of this annular (i.e. azimuthal)
electric current. These electrons/positrons emit incoherent
synchrotron radiation up to a cutoff frequency
\begin{eqnarray}
\nu_c & \equiv & \frac{3\gamma^3 c}{2\pi r_{lc}}=
\frac{3\gamma^3}{P_*}\nonumber\\
& \approx & 2\times 10^{18} \mbox{Hz}
\left(\frac{B_*}{10^{13}\mbox{G}}\right)^{3/2} \left(
\frac{P_*}{33\mbox{ms}} \right)^{-4}\ , \label{nuc}
\end{eqnarray}
which corresponds to hard X-rays. We emphasize once again that
this incoherent synchrotron radiation is due to the macroscopic
corotating motion of the separatrix electrons (positrons) around
the light cylinder. We would like to name this configuration {\em
`the pulsar synchrotron'}.

\section{Coherent radio emission}

What remains to be calculated is the total power radiated from
such an electron (positron) configuration. The power radiated by
an individual particle is equal to
\begin{equation}
L_e=\frac{2}{3}\frac{e^2 c}{r_{lc}^2}\gamma^4\sim
10^{-3}\mbox{erg}\
\mbox{s}^{-1}\left(\frac{B_*}{10^{13}\mbox{G}}\right)^2
\left(\frac{P_*}{33\mbox{ms}}\right)^{-6}\ ,
\end{equation}
with a frequency distribution (in erg/s/Hz)
\begin{equation}
F_\nu = 3.23\frac{e^2}{cP_*}(\nu P_*)^{1/3}
\end{equation}
at low frequencies $\nu \ll \nu_c$ (Jackson~1975). As seen in
fig.~(2), the annular volume where radiating electrons/positrons
attain their maximum Lorentz factor given by eq.~(\ref{gamma}) has
thickness and height comparable to $d$, and contains
\begin{equation}
N_e\sim 2\pi nr_{lc} d^2 \approx 8\times
10^{25}\left(\frac{B_*}{10^{13}\mbox{G}}\right)^{1/2}
\end{equation}
particles. The total incoherent power emitted by these particles
is probably too low to be observable. As we will now argue,
however, coherency effects may become very important in the radio
frequency range around 100~MHz.

It is well known that, in laboratory synchrotrons, particles
travel in bunches with macroscopic dimensions ($\sim$mm) along
their direction of motion, where the bunch length $\lambda$ is
determined by the geometrical and operational characteristics of
the synchrotron. It has been observed that, {\em at wavelengths
comparable to the bunch length (or the length of any structure in
the bunch), the radiation from multiple particles is in phase},
giving a radiated power proportional to $N^2$ times the power
radiated by an individual particle, where $N$ is the bunch
population (e.g. Venturin {\em et al.}~2005). This emission is
orders of magnitude stronger than the incoherent power at those
wavelengths, which is proportional to $N$. A practical application
of this effect is the generation of THz coherent synchrotron
radiation in synchrotrons with X-ray incoherent synchrotron
radiation (e.g. Sannibale {\em et al.}~2003). The pulsar
synchrotron is certainly very different from an earth synchrotron.
Nevertheless, we may also obtain a characteristic bunch length
$\lambda$ in the pulsar synchrotron as follows: separatrix
electrons/positrons stay in the annular volume of maximum Lorentz
factor at the tip of the dead zone for a path-length on the order
of
\begin{equation}
\lambda\sim d \gamma \sim \frac{r_{lc}}{2\gamma} \sim 3\times
10^2\mbox{cm} \left(\frac{B_*}{10^{13}\mbox{G}}\right)^{-1/2}
\left(\frac{P_*}{33\mbox{ms}}\right)^{2} \label{lambda}
\end{equation}
(fig.~2). If indeed particles emit coherently at the above
characteristic wavelength during the time they need to cross the
region of maximum Lorentz factor, the coherent radiation
corresponds to about 100~MHz. The annular radiating volume
consists of
\begin{equation}
\mu = \frac{2\pi r_{lc}}{\lambda} = 4\pi\gamma \sim 3\times 10^6
\left(\frac{B_*}{10^{13}\mbox{G}}\right)^{1/2}
\left(\frac{P_*}{33\mbox{ms}}\right)^{-1}
\end{equation}
regions of length $\lambda$, each one consisting of $N\equiv
N_e/\mu$ particles. The total luminosity of the coherent emission
in a frequency range on the order of $c/\lambda$ may be estimated
as the sum of the luminosities of the $\mu$ coherently emitting
regions
\begin{eqnarray}
L & \sim & 2\mu N^2 \nu F_\nu|_{\nu=c/\lambda}\nonumber \\
& \sim & 4\times 10^{28}\mbox{erg}\
\mbox{s}^{-1}\left(\frac{B_*}{10^{13}\mbox{G}}\right)^{7/6}
\left(\frac{P_*}{33\mbox{ms}}\right)^{-7/3} \label{L}
\end{eqnarray}
(the factor of 2 accounts for the contribution to the total radio
luminosity from both hemispheres of the dead zone separatrix).

This is just an order of magnitude calculation of the coherent
radio luminosity expected from the tip of the dead zone. More
detailed analysis will require the 3D numerical self-consistent
reconstruction of the electric and magnetic field geometry near
the tip of the dead zone together with the calculation of
separatrix particle orbits (work in progress). Nevertheless, our
estimate of the coherent radiation frequency is not far from the
observed peak of pulsar radio emission at a few hundred MHz.
Moreover, as we can see in fig.~(3), our result in eq.~(\ref{L})
is consistent with estimates of the total pulsar radio luminosity
obtained from the ATNF pulsar catalogue\footnote{Our estimates are
obtained as follows: for each catalogued pulsar with available
radio brightness data we calculate the ON brightness as $P_*/W$
times the average brightness, where $W$ is the observed pulse
width. The total radio luminosity is then calculated in the
context of the pulsar synchrotron model in which radiation is
emitted within $\pm 1/\gamma$ radians above and below the equator.
This is very different from the canonical model where radio
emission is produced in two beams originating in the polar caps.}
(Manchester {\em et al.}~2005). We therefore believe that our
scenario merits serious consideration as an alternative to the
radio pulsar phenomenon.

\section{Discussion}

The scenario that pulsar radio emission may be coming from the
light cylinder is not new. In the early days of pulsar astronomy,
people discussed the possibility that radio pulses may be due to
hot plasma corotating at discrete positions on the light cylinder
(e.g. Gold~1968, 1969; Bartel~1978; Cordes~1981; Ferguson~1981).
These models have since been abandoned. Our present model is
different in several respects (our plasma is cold in its rest
frame, the fundamental synchrotron radiation frequency is the
neutron star rotation frequency, pulses are due to the 3D
structure of the magnetosphere, etc.).

We would like to conclude this presentation with two final points
of interest. The first has to do with one of the few robust facts
about pulsar radio emission, namely the characteristic S-curve
polarization sweeps. In order for our model to survive as a valid
alternative to highly successful phenomenological models such as
the rotating vector model of Radhakrishnan \& Cooke~(1969), it has
to be able to explain this fact. This analysis requires knowledge
of the 3D structure of the pulsar magnetosphere obtained only
recently by Spitkovsky~(2006) and Kalapotharakos \&
Contopoulos~(2009). Let us, however, present here some preliminary
results, and let us introduce the three basic angles of the 3D
problem, namely the angle $\alpha$ between the rotation axis and
the stellar magnetic moment, the angle $i$ between the observer
line of sight and the rotation axis, and the observed polarization
angle $\psi$ with $\psi=0$ along the rotation equatorial plane.
Let us also introduce the angle $\xi$ between the rotation
equatorial plane and the direction connecting the central star
with the instantaneous position of the tip of the dead zone on the
light cylinder. In the context of our present pulsar synchrotron
model, radiation is observed whenever the instantaneous direction
of the annular current at the tip of the dead zone happens to lie
close to our line of sight. The instantaneous direction of
polarization $\psi$ lies along the direction of the instantaneous
radius of curvature which, we argue, must be close to $\xi$.
Therefore, $\psi\sim \tan^{-1}[\tan(\xi)\sin(i)]$. In fig.~(4) we
show the variation of $\xi$ with pulsar longitude as obtained in
the solution of Kalapotharakos \& Contopoulos~(2009) for
inclination $\alpha=60^o$. We also plot the resulting polarization
sweep for an observer inclination $i=30^o$. This preliminary
result is promising enough to justify the continuation of our
effort in paper~II of this series.

The second point of interest has to do with the equatorial return
current sheet beyond the light cylinder. One can directly check
that this is positively (negatively) charged, that is with a
surface charge {\em opposite} to that of the separatrix current
sheet inside the light cylinder. And yet, the electric current in
both current sheets flows in the same direction, that is towards
(away from) the star. Given the limitation on the available number
density of charge carriers imposed by the force-free approximation
(see discussion in \S~2), in order to supply the inward flowing
electrons (positrons) and the outward flowing positrons
(electrons), we need a source of pairs at the tip of the
separatrix at the position of the light cylinder. This may either
be supplied as a steady stream of pairs along the separatrix from
the surface of the star, or through pair creation by some yet to
be determined process at that position. Note that as these
particles move away from the tip of the dead zone, they either
move along the equator in a trajectory which becomes
asymptotically radial (i.e. with a radius of curvature much larger
than $r_{lc}$), or they move inwards where the corotation Lorentz
factor becomes smaller and smaller. In both cases, their
contribution to the total synchrotron radiation energy content is
expected to be much smaller than the radiation they emit as they
traverse the annular volume of thickness $d$ just inside the light
cylinder that we discussed in the previous section.

\section*{Acknowledgments}

We would like to thank G. Contopoulos, C. Efthymiopoulos, C.
Kalapotharakos and D. Kazanas  for stimulating discussions, and
the referee, A. Spitkovsky, for his constructive criticism.

\begin{figure}
\includegraphics[angle=0,scale=.50]{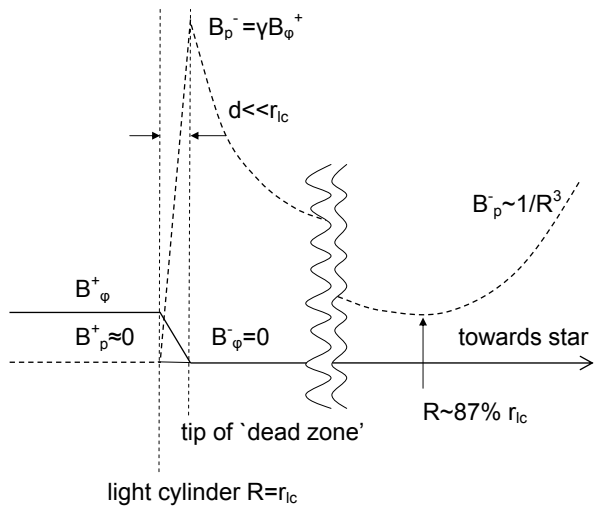}
  \caption{Schematic of the
  dependence of the poloidal (dashed line) and toroidal (solid line)
  magnetic field components
  on radial distance $R$ from the star
  along the equator near the tip of the dead zone.
  The poloidal field varies
  as $\propto 1/R^3$ near the star, reaches
  a minimum value at $87\%$ of the light cylinder distance, grows
  to a very large value at the tip of the dead zone as given by
  eq.~\ref{BpBphi}, and drops to zero outside.
  The toroidal field is zero inside the dead zone, and rises to some
  finite value outside. The jumps in $B_p$ and $B_\phi$ at the tip
  of the dead zone take place over some finite distance $d\ll r_{lc}$.}
\label{fig1}
\end{figure}

\begin{figure}
\includegraphics[angle=0,scale=.50]{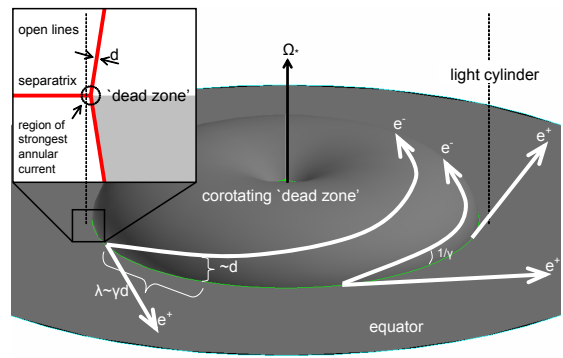}
  \caption{Schematic of electron and positron trajectories along the
  separatrix and the equatorial current sheet respectively
  for aligned magnetic and rotation axes, and detail of
  the tip of the dead zone. The light cylinder is denoted with
  a dashed line. $d\sim 10^{-3}$~cm is the separatrix
  thickness. $\lambda\sim 1$~m is
  the electron path length in the pulsar synchrotron.}
  \label{fig2}
\end{figure}

\begin{figure}
\includegraphics[angle=0,scale=.50]{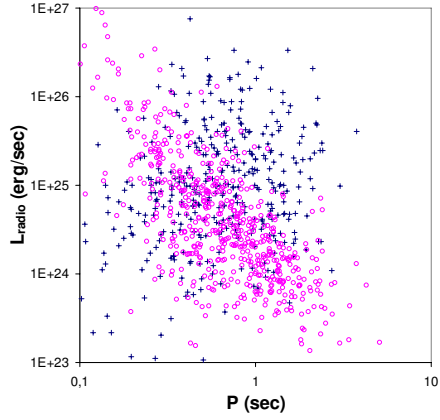}
  \caption{Pulsar radio luminosity as a function of the pulsar period.
  Open circles: calculated according to eq.~\ref{L}.
  Crosses: estimated from the observed emission at
  400~MHz adjusted in the context of the pulsar synchrotron model
  where radiation is emitted within $\pm 1/\gamma$ radians above and
  below the equator instead of in two beams originating
  in the polar caps. Data are taken from the ATNF pulsar catalogue. }
\label{fig3}
\end{figure}

\begin{figure}
\includegraphics[angle=0,scale=.50]{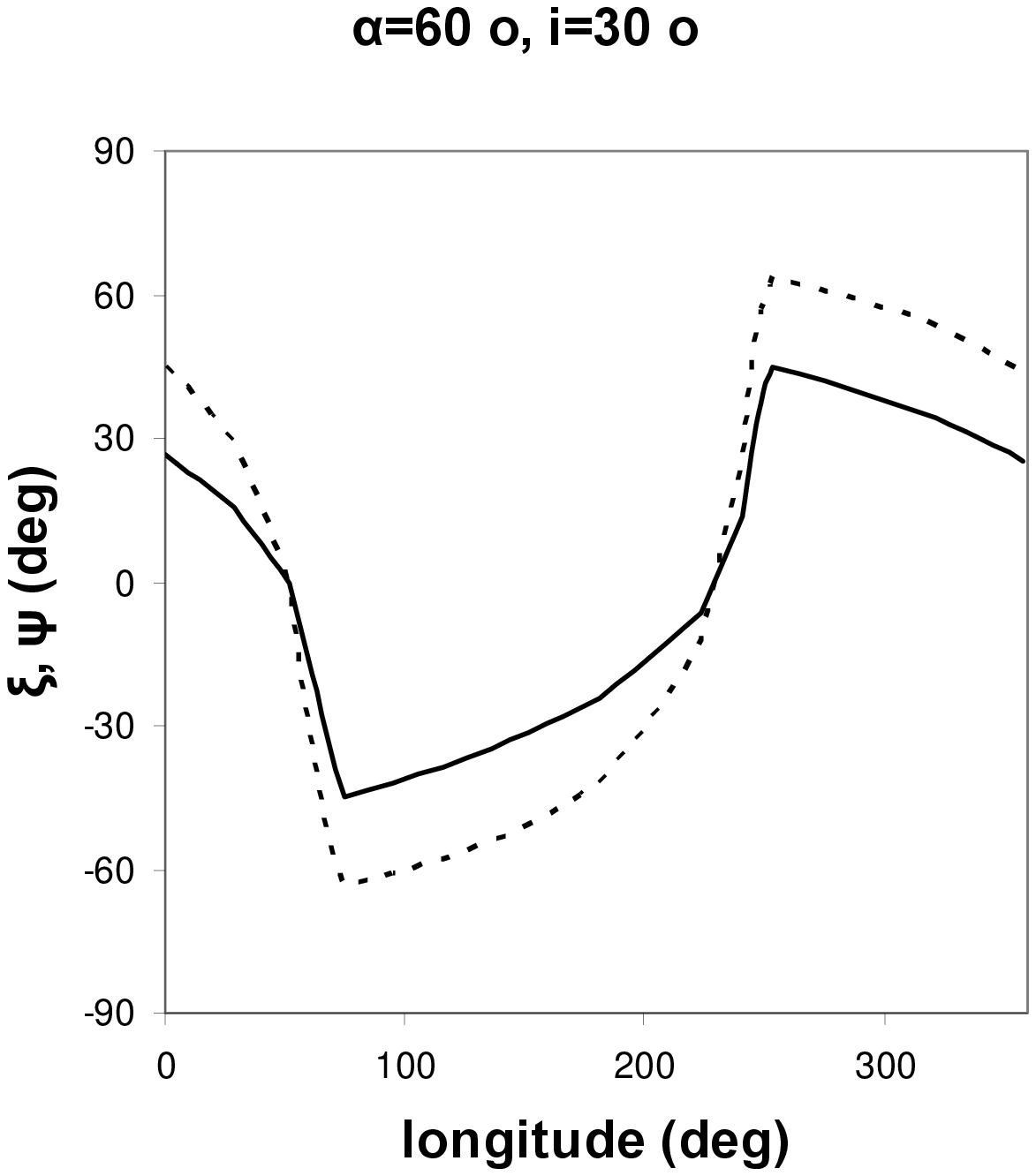}
  \caption{Distribution of $\xi$ (dotted line) and estimate of
    polarization angle $\psi$ (solid line)
    as functions of the pulsar longitude
    obtained from the numerical solution of Kalapotharakos \&
    Contopoulos~(2009) for $\alpha=60^o$ and $i=30^o$. }
\label{fig4}
\end{figure}

\label{lastpage}




\begin{thebibliography}{08}
\bibitem{A83} Arons, J. 1983, in 'Electron-Positron Pairs in
Astrophysics', eds. M. L. Burns, A. K. Harding \& R. Ramaty, 163
(New York: American Institute of Physics)
\bibitem{Ba77} Bartel, N. 1978, A\& A, 62, 393
\bibitem{CKF99} Contopoulos, I., Kazanas, D. \& Fendt, C. 1999, ApJ, 511, 351 (CKF)
\bibitem{C05} Contopoulos, I. 2005, A\& A, 442, 579
\bibitem{C81} Cordes, J. M. 1981, in `Pulsars: 13 years of
research on neutron stars', Proc. of the Symposium, 115 (Bonn:
Dordrecht)
\bibitem{F81} Ferguson, D. C. 1981, in `Pulsars: 13 years of
research on neutron stars', Proc. of the Symposium, 141 (Bonn:
Dordrecht)
\bibitem{G68} Gold, T. 1968, Nature, 218, 731
\bibitem{G69} Gold, T. 1969, Nature, 221, 25
\bibitem{GJ69} Goldreich, P. \& Julian, W. H. 1969, ApJ, 157, 869
\bibitem{G05} Gruzinov, A. 2005, Phys. Rev. Lett., 94, 021101
\bibitem{J75} Jackson, W. D. 1975, Classical Electrodynamics (New
York: Wiley)
\bibitem{K00} Kramer, M., Wex, N., \& Wielebinski, R. 2000, in
Pulsar Astronomy - 2000 and beyond, ASP Conf. Series, 202
\bibitem{K96} Kramer, M., Xilouris, K. M., Jessner, A.,
Lorimer, D. R., Wielebinski, R., \& Lyne, A. G. 1997, A\& A, 322,
846
\bibitem{ATNF} Manchester, R. N., Hobbs, G. B., Teoh, A. \&
Hobbs, M. 2005, AJ, 129, 1993
\bibitem{KC09} Kalapotharakos, C. \& Contopoulos, I. 2009, A\& A,
in press
\bibitem{RC69} Radhakrishnan, V. \& Cooke, D. J. 1969, AL, 3, 225
\bibitem{S03} Sannibale, F., Byrd, J. M. Loftsdottir, A., Martin,
M. C., \& Venturini, M. 2003, in Proc. of the 2003 Particle
Accelerator Conference, IEEE
\bibitem{S06} Spitkovsky, A. 2006, ApJ, 648, L51
\bibitem{T06} Timokhin, A. N., 2006, MNRAS, 368, 1055
\bibitem{U03} Uzdensky, A. D. 2003, ApJ, 598, 446
\bibitem{V05} Venturini, M.,Warnock, R., Ruth, R., \&
Ellison, J. A. 2005, PhysRevSTAB, 8, 014202
\end{thebibliography}
\end{document}